\documentclass[preprint, 12pt,nofootinbib]{revtex4-1}
\usepackage[T1]{fontenc} 

\usepackage{graphicx}
\usepackage{cancel}
\usepackage{amssymb}
\usepackage{textcomp}
\usepackage{amsmath}
\usepackage{bm}
\usepackage{times}
\usepackage{epsfig}
\usepackage{color}

\def\beq{\begin{equation}}
\def\eeq{\end{equation}}
\def\be{\begin{equation}}
\def\ee{\end{equation}}
\def\bea{\begin{eqnarray}}
\def\eea{\end{eqnarray}}
\def\to{\rightarrow}

\begin{document}
\title{Searching Supersymmetric Leptonic Partner at the CERN LHC}
\author{Jiwei Ke$^{1}$, Ming-Xing Luo$^{1}$, Lian-You Shan$^{2}$, Kai Wang$^{1}$\footnote{Corresponding Author: wangkai1@zju.edu.cn} and Liucheng Wang$^{1}$}
\affiliation{$^{1}$~Zhejiang Institute of Modern Physics and Department of Physics, Zhejiang University, Hangzhou, Zhejiang 310027, CHINA\\
$^{2}$~Institute of High Energy Physics, Chinese Academy of Sciences, Beijing 100049, CHINA}

\begin{abstract}
Motivated by the observed excess of the di-photon signal in Higgs 
searches, $\sigma_{\gamma\gamma}/\sigma_{\rm SM}\simeq 1.5$, we argue that models with enhanced $\Gamma(h\to \gamma\gamma)$ alone are the most favorable scenarios when the latest LHC/Tevatron results are all taken into account. We study 
the phenomenology of a supersymmetric scenario of light stau first proposed by \textit{Carena et. al. \cite{carlos}} that predicts a 125~GeV SM-like Higgs boson with enhanced diphoton decay through light stau loops. Since it is extremely challenging to search the Drell-Yan stau pair at the LHC due to the small production rate, we focus on the parameter space with enhanced production of inclusive stau pairs, 
in particular, via $b\bar{b}$ fusion or gaugino pairs. We study its phenomenology in both pure leptonic tau $\tau^\pm_\ell$ channels and hadronic tau tagged $\tau_h$ channels. We find the same-sign dilepton from $\tilde{\chi}^\pm_1 \tilde{\chi}^0_2 \to \tau^\pm_\ell \tau^\pm_\ell+X$ may significantly improve the discovery potential with even 7--8 TeV LHC of ${\cal O}(30~\text{fb}^{-1})$ data. In the case of hadronic tau pair, we use the final state $j+\tau_h \tau_h +\cancel{E}_T$ to search
and find that even with the most optimistic region of $M_2\sim 200-300$~GeV, it requires at least 50~fb$^{-1}$ data of 14 TeV LHC to reach a significance of 3.5 $\sigma$. Therefore, we conclude it is
difficult to claim discovery only through hadronic tau based on the data by the 2012 shut-down. 5 $\sigma$ reach for our most optimistic region then requires 100~fb$^{-1}$ data  with 14 TeV running. 
\end{abstract}

\maketitle
\flushbottom

\section{Introduction}

The ATLAS and CMS experiments at the CERN Large Hadron Collider (LHC) have both reported a combined 5~$\sigma$ discovery of Higgs-like boson in the two channels of highest resolution, the di-photon ($gg\to h\to \gamma\gamma$) and four-lepton ($gg\to h\to ZZ^{*}\to 4 \ell$) with the data of integrated luminosity of 5~fb$^{-1}$ at 7 TeV plus 5~fb$^{-1}$ at 8 TeV running \cite{today}. Both di-photon and four-lepton have shown hints of resonance with reconstructed invariant mass around $m_{\gamma\gamma}\sim m_{4\ell}\sim 125$~GeV. 
Excess in the pure leptonic $WW^*$ was also confirmed later in the consistent range of the four-lepton from $ZZ^*$ \cite{ww}.
Precision measurement of properties of such resonance will play important role to determine whether it is the standard model (SM) Higgs. Distinct signals of di-photon and four-lepton modes clearly show that there exists a spin zero or spin two resonance that couples to weak gauge bosons. At hadron colliders, the $s$-channel resonance can be produced via light quark annihilation or gluon fusion. Since the scalar couples to the left-handed and right-handed fermions, a light quark mass is then proportional to the coupling between this scalar and light quark. If the resonance is a scalar, gluon fusion production becomes the only option. The gluon fusion ($gg\to h$) production and the di-photon decay $h\to \gamma\gamma$ are  both loop induced  which are sensitive to physics involved in the loops. Measurements of such channels provide ways to probe new physics of higher scale at the same time. 

Latest datas from both ATLAS and CMS have shown that the measured di-photon
rate is significantly larger than the SM prediction while the four-lepton rate is about the same as the SM prediction at 125 GeV. Results from the CMS collaboration are 
\beq
\frac{\sigma(gg\to h\to \gamma\gamma)}{\sigma(gg\to h_{\rm SM}\to \gamma\gamma)} \simeq 1.56\pm 0.43
\eeq
where $h$ is the higgs-like resonance \cite{today}. In the four-lepton channel, the expected SM significance of the SM Higgs is 3.8~$\sigma$ while the observed value is 3.2~$\sigma$. The ATLAS
collaboration have reported 4.5~$\sigma$ significance in di-photon while the SM Higgs expectation is
2.4~$\sigma$ and results in four-lepton channel showed an observed 3.4~$\sigma$ significance versus a 2.6~$\sigma$ expectation.
The other channels with  $h\to b\bar{b}$  and $h\to \tau^{+}\tau^{-}$ are more challenging experimentally than the previous channels.  The ATLAS collaboration sees no excess in the above channels while CMS collaboration reported about $1~\sigma$ in the channels. On the other hand, the CDF  and  $D\cancel{0}$ collaborations at Tevatron have both reported excess of $W b\bar{b}$ in the light Higgs mass range of 115-130~GeV which is consistent with the LHC findings of 124-126~GeV \cite{tevatron}. The excess at Tevatron is crucial to test various models. 

For given scalar mass $m_{h}$, the gluon fusion $s$-channel resonance $h$ production with decaying into diphoton, $gg\to h\to \gamma\gamma$ is approximately proportional to 
\beq
\frac{\Gamma(h\to gg)\Gamma(h\to \gamma\gamma)}{\Gamma_{\rm total}(h\to {\rm all})},
\eeq
thus there only exist three categories to enhance the di-photon: enhanced $h\to gg$, enhanced $h\to \gamma\gamma$ or the reduced total width $\Gamma_{\rm total}$. One scenario is through increased $\Gamma(h\to gg)$ with total width increasing at slower rate. The second scenario is through increased $\Gamma(h\to \gamma\gamma)$. And the third one is the reduced total width, for instance, the reduced $\Gamma(h\to b\bar{b})$ through mixing.  
Example of the first category is models with fourth generation fermions\cite{4th} or enhanced Yukawa models, like Bosonic TechniColor \cite{luohui} and Type-I Two-Higgs-Doublet Model (2HDM), in
which there exists a Higgs-like doublet scalar but at the same time,
there also exists fermiophobic sector that contribute to the $W$/$Z$ mass. Within SM, $h\to \gamma\gamma$ decay involves two types of contribution of $W$ loop and top quark loop with opposite signs\cite{luohui}. The SM $W$-loop dominates $h\to \gamma\gamma$ partial width. The enhancement in $h\to gg$ inevitably enhances the cancellation in the $h\to \gamma\gamma$. The enhanced Yukawa model predicts reduced couplings with the weak gauge bosons $W$ and $Z$ which leads to significant reduction in the associated production of Higgs $Wh$ or $Zh$ with enlarged $h\to b\bar{b}$ decay branching fraction which may fit the Tevatron observations. However, for $m_h=125~$GeV, the fermionic
decays $h\to b\bar{b}$, $h\to\tau^+\tau^-$ still dominate the Higgs total width. Enhancement due to top quark Yukawa is then cancelled by 
the enhanced total width from fermionic decays. The diphoton partial
width must increase in order to achieve the enhanced diphoton rate
while this corresponds to an un-natural region that the $W$-mass $m_W$ mostly arises from the fermiophobic sector. 
The model with fourth generation fermions predicts similar rate of associated production $Wh$ as the SM Higgs prediction. However, in order to explain the measurement of $gg\to h\to ZZ^{*}\to 4\ell$, it usually requires significant reduction of all known decay modes in both $h\to b\bar{b}$ and $h\to ZZ^{*}$ by introducing new decay channel to fourth generation neutrinos $h\to N\bar{N}$.  In the third category, models with reduced total width by reducing $h\to b\bar{b}$ and $h\to WW^{*}$ through mixing also predicts suppressed $Wb\bar{b}$ at Tevatron. Therefore, the over 2~$\sigma$ excess at Tevatron of $W b\bar{b}$ does not favor the first category or the third category of models.  Consequently, models with enhanced $\Gamma(h\to \gamma\gamma)$ \cite{ian} become the most favorable models for combined measurements at both
LHC and Tevatron.  

In order to achieve the enhanced $h\to \gamma\gamma$ without effecting $gg\to h$, the model requires non-colored charged states that couple to the Higgs boson. In the extension of SM, there are several candidates of this kind,
for instance, $W^{\prime}$, charged Higgs, heavy leptons or leptonic partners in supersymmetric theory \cite{ian}.  
We study the light stau in supersymmetric standard models to illustrate models with the enhanced $h\to \gamma\gamma$. The scenario was first proposed by \cite{carlos} and further studied in \cite{liantao}. However, with only electroweak production, light stau is extremely challenging to search for at the LHC \cite{liantao}. In this paper, we focus on the the benchmark scenario in MSSM with enhancement in stau production and at the same time with the prediction of 125~GeV Higgs that has enhanced $\gamma\gamma$ rate and predictions consistent with other observations. 
In particular, we emphasize how the light stau production can be significantly enhanced from not-very-heavy $M_A$ 
and inclusive stau pairs from gaugino productions.  Before we conclude, we discuss the kinematic feature and searching potential at the LHC for such enhanced inclusive stau pair production. 
 
\section{Light staus and its phenomenology}

Low energy supersymmetry has been the most elegant solution to the gauge hierarchy problem
in the last three decades. In decoupling region of MSSM, the lighter CP even 
neutral scalar $h$ behaves as a SM-like Higgs boson and we assume the discovered 
resonance is the $h$ boson. The strongest motivation for supersymmetry
is the cancellation of quadratic divergence, in particular, the top quark 
contribution due to large top Yukawa. At the same time, the squarks
also significantly modify the Higgs production via gluon fusion and
Higgs decaying into di-photon  \cite{ggh}. Since the data set for the 
four-lepton final states are about the same as the SM Higgs prediction, we 
argue the gluon fusion production should not be changed significantly 
by the effects due to squarks. 

The general sfermion mass matrix is       
\begin{equation}
{\cal M}^2_{\tilde{f}} = \left(
  \begin{array}{cc} m_{\tilde{f}_L}^2 + m_f^2 + D_L^f & m_f \, \tilde{A}_f \\
                    m_f \tilde{A}_f & m_{\tilde{f}_R}^2 + m_f^2+ D_R^f 
\end{array} \right)
\label{matrix}
\end{equation}
where the off--diagonal entries are $\tilde{A}_t= A_t-\mu\cot\beta$ for top squark with 
$\tan\beta$ the ratio of the vacuum expectation values of the 
two--Higgs fields which break the electroweak symmetry, $A_t$  
the trilinear squark coupling which breaks the $R$-symmetry, and $\mu$ Higgsino mass parameter, 
respectively. $m_{\tilde{f}_L}$ and $m_{\tilde{f}_R}$ are the left-- and 
right--handed soft--SUSY breaking sfermion masses. The $D$ terms, in units of $M_Z^2 \cos 
2\beta$ are given in terms of  the weak isospin and electric charge of 
the squark by:  $D_L^f= I^{3}_f -e_f \sin^2\theta_W$ and $D_R^f= e_f\sin^2
\theta_W$. The leading contribution is from top squark (stop) and bottom squark in the limit of large $\tan\beta$. We take $\tan\beta=30$ in the following discussion after taking into account the 
constraint from $B_{s}\to \mu^{+}\mu^{-}$. So the effect is mostly due to top squark.
To calculate the stop effect, one obtains the physical states $\tilde{t}_{1}$ and $\tilde{t}_{2}$ 
from the mass matrix in Eq.\ref{matrix}. The couplings of the physical squark pairs to the Higgs boson $h$, normalized to $2M_Z^2(\sqrt{2}G_F)^{1/2}$, are
\begin{eqnarray}
g_{h \tilde{f}_1 \tilde{f}_1 } &=& - \cos 2\beta \left[ 
I_f^3 \cos^2 \theta_{\tilde{f}} - e_f \sin^2 \theta_W \cos 2
\theta_{\tilde{f}} \right] 
- \frac{m_f^2}{M_Z^2} + \frac{1}{2} \sin 2\theta_{\tilde{f}} 
\frac{m_f \tilde{A}_f } {M_Z^2} \nonumber \\
g_{h \tilde{f}_2 \tilde{f}_2 } &=& - \cos 2\beta \left[ 
I_f^3 \sin^2 \theta_{\tilde{f}} - e_f \sin^2 \theta_W \cos 2
\theta_{\tilde{f}} \right] 
- \frac{m_f^2}{M_Z^2} - \frac{1}{2} \sin 2\theta_{\tilde{f}} 
\frac{m_f \tilde{A}_f} {M_Z^2}
\label{coupling}
\end{eqnarray}
The gluon fusion production is proportional to the decay partial 
width of $h\to gg$ which is given by
\beq
\Gamma (h \to gg) = \frac{G_F \alpha^2_s 
M_{h}^3}{64  \sqrt{2} \pi^3 } \left| \, 
\sum_Q A_Q (\tau_Q)  +  \sum_{\tilde{Q}} g_{h \tilde{Q} \tilde{Q}} \, 
\frac{M_Z^2} {m_{\tilde{Q}}^2} \, A_{\tilde{Q}} (\tau_{\tilde{Q}}) \, 
\right|^2
\eeq
where the scaling variable $\tau_i$ is defined as $\tau_i = M_{h}^2/
4m_i^2$ with $m_i$ the mass of the loop particle, and amplitudes 
$A_i$ are  
\begin{eqnarray}
A_{Q}(\tau) &=& - 2 [\tau+(\tau-1)f(\tau)]/\tau^2 \nonumber \\
A_{\tilde{Q}} (\tau) &=& [\tau -f(\tau)]/\tau^2~.
\end{eqnarray}
Function $f(\tau)$ is
\begin{equation}
f(\tau) = \left\{ \begin{array}{ll} 
{\rm arcsin}^2 \sqrt{\tau} & \tau \leq 1 \\
-\frac{1}{4} \left[ \log \frac{1 + \sqrt{1-\tau^{-1} } }
{1 - \sqrt{1-\tau^{-1}} } - i \pi \right]^2 \ \ \ & \tau >1 
\end{array} \right. 
\end{equation} 
Large $\tilde{A}_{t}$ significantly enhances $g_{h\tilde{t}_{1}\tilde{t}_{1}}$, the coupling of squark pairs to $h$ which results in large cancellation in the $\Gamma(h\to gg)$ for light $m_{\tilde{t}_{1}}$
of $\cal O$(200~GeV) \cite{ggh}. On the other hand, the $\tilde{A}_{t}$ and $m_{\tilde{t}_{1}}$ are also constrained by the Higgs mass $m_{h}$.
  
At tree-level, the mass of $h$ is bounded as $m_{h}^{2}\leq m_{Z}^{2}\cos^{2}2\beta$ \cite{Inoue:1982ej,Flores:1982pr}. The dominating loop contribution comes from the top/stop
sector. Up to 1-loop precision, the mass of the $h$ boson is
given by the formula \cite{Carena:1995bx}: 
\begin{equation}
m_{h}^{2}\simeq m_{Z}^{2}\cos^{2}2\beta+\frac{3m_{t}^{4}}{4\pi^{2}v^{2}}\left[\log\frac{M_{\mathrm{SUSY}}^{2}}{m_{t}^{2}}+\frac{\tilde{A}_{t}^{2}}{M_{\mathrm{SUSY}}^{2}}\left(1-\frac{\tilde{A}_{t}^{2}}{12M_{\mathrm{SUSY}}^{2}}\right)\right],\label{eq:higgs mass}
\end{equation}
where $m_{t}=172.9$ GeV being the top quark mass, $v=174$ GeV being
EWSB VEV, $M_{\mathrm{SUSY}}^{2}=m_{\tilde{t}_{1}}m_{\tilde{t}_{2}}$
being the averaged stop mass square \footnote{The
above formula is valid only for small splitting between two stop masses
and no thresholds effects \cite{carlos}. Moreover, Eq. \ref{eq:higgs mass} does
not include the sbottom and stau contributions, which could be important
for large values of $\tan\beta$.}. In this paper, we use \textsf{FeynHiggs} \cite{Heinemeyer:1998yj},
which has taken account of all the contributions at the two-loop level, to calculate the masses of the Higgs bosons. 

To realize a 125 GeV Higgs boson in the MSSM, there have
been many successful attempts \cite{carlos,Hall:2011aa,Baer:2011ab,Li:2011ab,Heinemeyer:2011aa,Arbey:2011ab,Arbey:2011aa,Kang:2012tn,Desai:2012qy,Cao:2012fz,Christensen:2012ei}. Generally speaking, loop contribution
to $m_{h}$ need to be significant. Certain amount of fine-tunings
are necessary since the stop masses and the mixing parameter $\tilde{A}_{t}$
must be judiciously chosen. In order to have a feeling on the fine-tuning,
we have a close look at Eq. \ref{eq:higgs mass}. At first sight,
one can choose $M_{\mathrm{SUSY}}^{2}/m_{t}^{2}\gg1$ in Eq. \ref{eq:higgs mass}
to enhance the loop contribution. But this set of parameters will
result in the split SUSY which relaxes the assumption about naturalness.
There is another milder way to enhance loop contribution 
by choosing $M_{\mathrm{SUSY}}^{2}/m_{t}^{2}>1$ and $\tilde{A}_{t}^{2}/M_{\mathrm{SUSY}}^{2}>1$
in Eq. \ref{eq:higgs mass}. Namely, the stop masses are of order
of several hundred GeV to several TeV as well as a large mixing parameter
$\tilde{A}_{t}$. In order to minimize the stop effect in 
the gluon fusion production, we choose large $m_{\tilde{t}}$ and the relevant input as 
\beq
M_{\tilde{q}}=1.5~\text{TeV}, M_{\tilde{u}}=1.5~\text{TeV}, M_{\tilde{d}}=2~\text{TeV}, A_{t}=A_{b}=2.5~\text{TeV}~.
\eeq 

As discussed in the introduction, we focus on the light stau contribution to enhance $\Gamma(h\to \gamma\gamma)$ in this paper.  
Similar to the stop states, large $A_{\tau}$ also induces large splitting in the stau mass eigenstates
as in Eq. \ref{matrix}. The di-photon enhancement amplitude 
is given in \cite{liantao} as 
\beq
\Delta A_{\gamma\gamma} \propto -\frac{(\mu\tan\beta)^2 m^2_\tau}{3[M^2_{L_3} m^2_{e_3}-m^2_\tau (\mu\tan\beta)^2]}
\eeq 
which clearly indicates large $\mu\tan\beta$ corresponds to the enhanced diphoton rate. 

However, a extremely sensitive measurement on $\tan\beta$ is through  
$B_{s}\rightarrow\mu^{+}\mu^{-}$ and $B\rightarrow X_{s}\gamma$. Recently, LHCb Collaboration
reported a new limit $\mathrm{Br}(B_{s}\rightarrow\mu^{+}\mu^{-})<4.5\times10^{-9}$ \cite{TALK},
which yields a new bound on $\tan\beta$.  $\mathrm{Br}(B_{s}\rightarrow\mu^{+}\mu^{-})$
is expected to be proportional to $\tan^{6}\beta$ in the MSSM \cite{Huang:2000sm}. A
SUSY model with a large $\tan\beta$ is now disfavored by this constraint.
Moreover, the Belle measurement on $b\rightarrow s\gamma$ is $\mathrm{Br}(B\rightarrow X_{s}\gamma)=(3.55\pm0.24)\times10^{-4}$ \cite{Limosani:2009qg}.
In this paper, we require that $\left|\mathrm{Br}(B\rightarrow X_{s}\gamma)-3.55\times10^{-4}\right|<1.0\times10^{-4}$
to constrain the MSSM parameters. 
In the following discussion, we fix the $\tan\beta$ value to be 30 and tune the $\mu$-term to a larger value to enhance the stau effects in the $h\to \gamma\gamma$. 
Furthermore, in supersymmetric models, the charged Higgs contribution to $b\to s$ transition is often cancelled by the stop/Higgsino contribution which is proportional to $A_{t}\mu$. Large $A_{t}$ is a requirement by the Higgs mass above 120~GeV and large $\mu$ is preferred
to induce the large splitting in stau states. Therefore, constraints on $b\to s\gamma$ is easily satisfied by the model choices.
We choose the SUSY input as \footnote{We focus on how stau production at LHC can be enhanced and therefore, we only choose one parameter point in the typical stau scenarios to illustrate the qualitative feature. Instead, we vary the two parameters $M_{A}$ and $M_{2}$ that strongly affect the phenomenological
results. }
\beq
M_{\tilde{\tau}_{L}}=M_{\tilde{\tau}_{R}}=350~\text{GeV},A_{\tau}=2.5~\text{TeV}, \mu=2.15~\text{TeV}, \tan\beta=30~,
\eeq
which results in
\beq
m_{\tilde{\tau}_{1}}\simeq 120~\text{GeV}.
\eeq

\begin{figure}
\begin{center}

\includegraphics[scale=1,width=8cm]{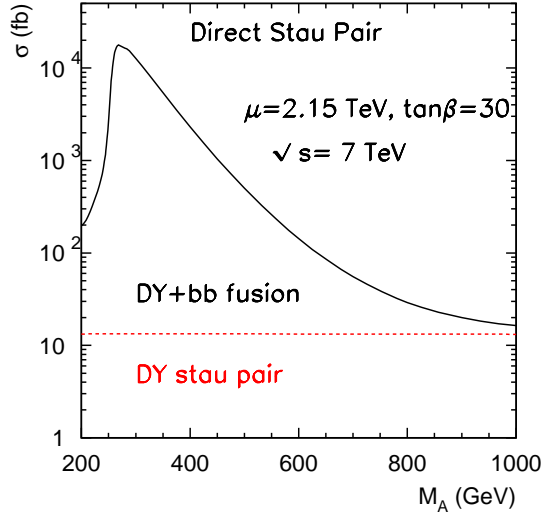}
\end{center}

\caption{Direct stau pair production at 7 TeV LHC, DY production plus $b\bar{b}$ fusion.} 
\label{marate}
\end{figure}

Slepton searches at the hadron collider have been very challenging. The Drell-Yan(DY) production of 
light stau pair is only at $\cal O$(10~fb) for extremely light stau as $m_{\tilde{\tau}_{1}}=120~\text{GeV}$.
However, in supersymmetric models, direct stau pair production may be dominated by the bottom fusion 
\beq
b \bar{b}\to h,H,A \to \tilde{\tau}^{+}_{1}\tilde{\tau}^{-}_{1},
\eeq
while
the $gg\to \tilde{\tau}^{+}_{1}\tilde{\tau}^{-}_{1}$ is sub-leading \cite{maike}. 
Figure \ref{marate} shows the production of direct stau pair including DY and the $b\bar{b}$ fusion at 7 TeV LHC.
In principle, in the limit of large $\mu\tan\beta$, coupling between $b\bar{b}$ to $h,H$ must be
corrected. Since supersymmetric contribution to the SM fermion mass breaks $R$-symmetry
and $PQ$-symmetry, the contribution must be proportional to the gaugino mass as well as
the $\mu$-term. In the decoupling limit of $M_A \gg m_Z$, the $h$ coupling is approaching
SM value at the tree level but the $H b\bar{b}$ coupling of
\beq
\frac{\cos\alpha}{\cos\beta}
\eeq
can be significantly enhanced by the large $\tan\beta$ which lead to
large enhancement in $b\bar{b}$ fusion produced stau pair in Fig. \ref{marate}.  
In addition, the bottom or tau Yukawa couplings can be reduced due to 
supersymmetric correction when $\mu\tan\beta$ is large. However, the supersymmetric correction to
couplings to $H$ is proportional to $\sin\alpha$ which is suppressed in the decoupling limit.
Therefore, we don't consider the supersymmetric correction of $H b\bar{b}$ coupling.
The solid line shows the resonant effect in the low $M_{A}$ region. On the other hand, $M_{A}$ 
is also constrained by supersymmetric contribution to $B_{s}\to \mu^{+}\mu^{-}$ and $b\to s\gamma$.
For fixed $\tan\beta=30$, we find 
\beq
M_{A}> 600~\text{GeV}.
\eeq
Then the largest possible rate of direct stau pair production for fixed $\tan\beta$ and $\mu$ is about 100 fb.

With conserved $R$-parity, the thermal relic abundance of the lightest neutralino (LSP) 
can often be identified with cosmic dark matter, consistent with the current cosmological observations $\Omega_{\mathrm{DM}}h^{2}=0.1120\pm0.0056$ \cite{Komatsu:2010fb}. Pure bino is a SM singlet
and its annihilation through $t$-channel scalar is an important channel. In the light stau scenarios,
bino can annihilate into tau pairs $\tilde{\chi}^{0}_{1} \tilde{\chi}^{0}_{1}\to \tau^{+}\tau^{-}$. 
To reproduce the required thermal relic abundance, we use \textsf{MicrOMEGAs} \cite{Belanger:2006is} and find 
\beq
M_{1} = 85~\text{GeV}, M_{2} > 125~\text{GeV}.
\eeq
In addition, we also checked the other constraints such as muon anomalous magnetic moment. The discrepancy between experiments and the SM calculation is $a_{\mu}^{\mathrm{EXP}}-a_{\mu}^{\mathrm{SM}}=(25.5\pm8.2)\times10^{-10}$ \cite{Davier:2009zi}.
In this paper, we require that the SUSY contribution to g-2 should
be smaller than this deviation.  Constraint from EWSB, such as $\triangle\rho^{\mathrm{SUSY}}<10^{-3}$ \cite{Altarelli:1998xf} is also included.

Since the lightest stau $\tilde{\tau}^{\pm}_{1}$ is the next-to-lightest supersymmetric particle (NLSP), all other supersymmetric particles can cascade decay into the on-shell stau $\tilde{\tau}^{\pm}_{1}$ final states if the phase space allows. When the two body modes dominate the gaugino decays,
\beq
\text{Br}(\tilde{\chi}^{-}_{1}\to \tilde{\tau}^{-}_{1} \bar{\nu}_{\tau})=100\%, 
\text{Br}(\tilde{\chi}^{0}_{2}\to \tilde{\tau}^{\pm} \tau^{\mp})=100\%
\eeq
with 50\% of $\tilde{\chi}^{0}_{2}$  decay into $\tau^{+}$ and the other 50\% to $\tau^{-}$.
The gaugino pairs completely turn into multi-stau final states as 
\beq
pp \to \tilde{\chi}^{\pm}_{1}\tilde{\chi}^{0}_{2}\to \tilde{\tau}^{\pm}_{1} \tilde{\tau}^{\mp}_{1} \tau^{\pm}\nu_{\tau}; \tilde{\chi}^{+}_{1}\tilde{\chi}^{-}_{1}\to \tilde{\tau}^{+}_{1} \tilde{\tau}^{-}_{1} \nu_{\tau}\bar{\nu}_{\tau},
\eeq
and generate the multi-tau final states plus large missing transverse energy $\cancel{E}_{T}$.
In the limit when $M_{2}\to 125$~GeV, $\tilde{\chi}^{\pm}_{1}$ and $\tilde{\chi}^{0}_{2}$ are nearly
degenerate with $\tilde{\tau}^{\pm}_{1}$. The additional $\tau^{\pm}$ associated with the $\tilde{\tau}^{\pm}_{1}$ then becomes extremely soft and cannot pass the basic detector cuts. Therefore, the final
states are indistinguishable from the stau pair production final states. 
One particular interesting final state is when 
\beq
pp \to \tilde{\chi}^{\pm}_{1}\tilde{\chi}^{0}_{2}\to \tilde{\tau}^{\pm}_{1} \tilde{\tau}^{\pm}_{1} \tau^{\mp}\nu_{\tau},
\eeq
which may fall into the same-sign di-lepton plus large $\cancel{E}_{T}$ search.
In Fig. \ref{rate}, we plot the production rate for gaugino pairs for varying $M_{2}$ at 7 TeV LHC.
\begin{figure}
\begin{center}
\includegraphics[scale=1,width=8cm]{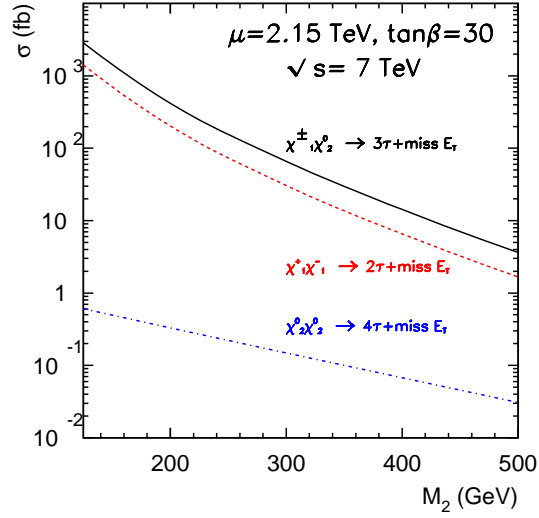}
\end{center}
\caption{Gaugino Production Rate with $\tilde{\chi}^{\pm}_{1}\to \tilde{\tau}^{\pm}_{1} \nu_{\tau}$ and $\tilde{\chi}^{0}_{2}\to\tilde{\tau}^{\pm}_{1}\tau^{\mp}$ at 7 TeV LHC } 
\label{rate}
\end{figure}
For small $M_{2}$, the production is one order higher than the direct stau pair production with $M_{A}=600$~GeV and completely dominates the stau pair final states.

Finally, we summarize the benchmark scenario with fixed 
\begin{eqnarray}
& M_{1}=85~\text{GeV}, M_{3}= 1.2~\text{TeV},\nonumber\\
 & \tan\beta = 30 ,  \mu= 2.15~\text{TeV}\nonumber\\
& M_{\tilde{Q}^{1,2,3}_{L}}=M_{\tilde{u}^{1,2,3}_{R}}=1.5~\text{TeV}, M_{\tilde{d}^{1,2,3}_{R}}=2~\text{TeV}\nonumber\\
& M_{\tilde{\ell}_{L}^{1,2}}=1.5~\text{TeV}, M_{\tilde{e}_{R}^{1,2}}=2~\text{TeV}\nonumber\\
& A_{t}=A_{b}=A_{\tau}=2.5~\text{TeV}, M_{\tilde{\tau}_{L,R}}=350~\text{GeV}.
\end{eqnarray}
We vary the $M_A$ and $M_2$ from
\beq
M_A : 600-900~\text{GeV~~~~and~~} M_2: 125-500 \text{GeV},
\eeq
to study how the two variables can change the inclusive stau pair production rate
at the LHC.

The benchmark scenario predicts the MSSM spectrum as
\beq
m_{h} = 124.05~\text{GeV}, m_{\tilde{\tau}_{1}}=120.0~\text{GeV}, m_{\tilde{\chi}^{0}_{1}}=84.91~\text{GeV}~.
\eeq
The predictions of di-photon, four-lepton and $b\bar{b}$ channels are 
\beq
{\sigma(gg\to h\to \gamma\gamma)\over \sigma_{SM}} = 1.52,  {\sigma(gg\to h\to ZZ^{*})\over \sigma_{SM}} =1.04,  {\sigma(W h\to b\bar{b})\over \sigma_{SM}}= 0.92
\eeq
where $\sigma_{SM}$ is the corresponding predictions of the SM Higgs boson. 

\section{Searching staus at the LHC}

At Hadron Colliders, light stau pair can be directly produced via Drell-Yan, $b\bar{b}$ fusion and gluon fusion. It turns out that the $b\bar{b}$ fusion is the leading production for $M_{A}$ of 
a few hundreds GeV. Shown in Fig.\ref{marate}, at the benchmark point with $M_{A}=600$~GeV,  
the Drell-Yan plus $b\bar{b}$ fusion production rate at 7 TeV LHC is about 100~fb. In addition, 
stau pair from gaugino pair productions can be much larger than the direct stau production. 
At 14 TeV LHC, the production rates are given in Fig. \ref{rate}
\begin{figure}
\begin{center}
\includegraphics[scale=1,width=8cm]{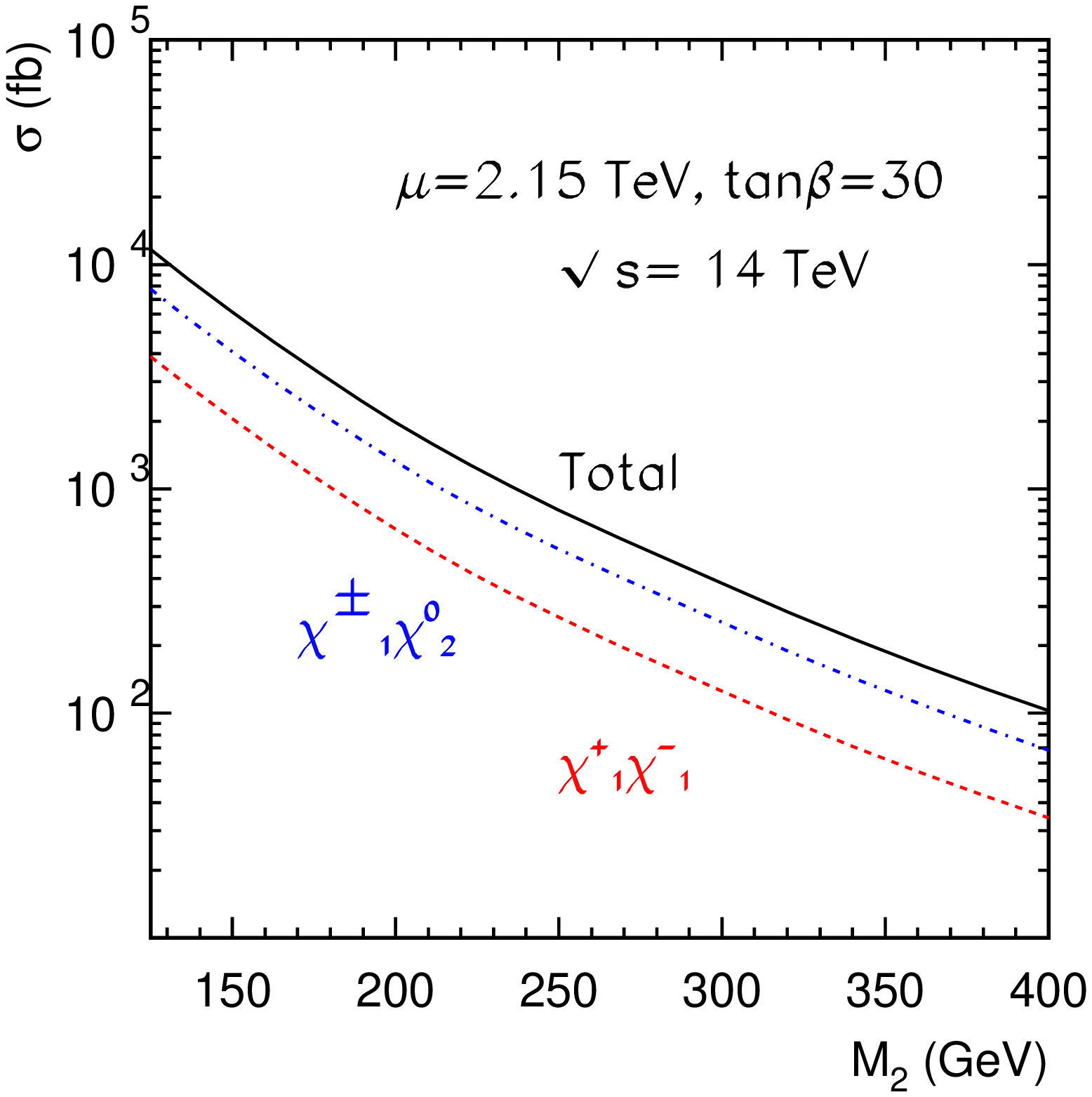}
\includegraphics[scale=1,width=8cm]{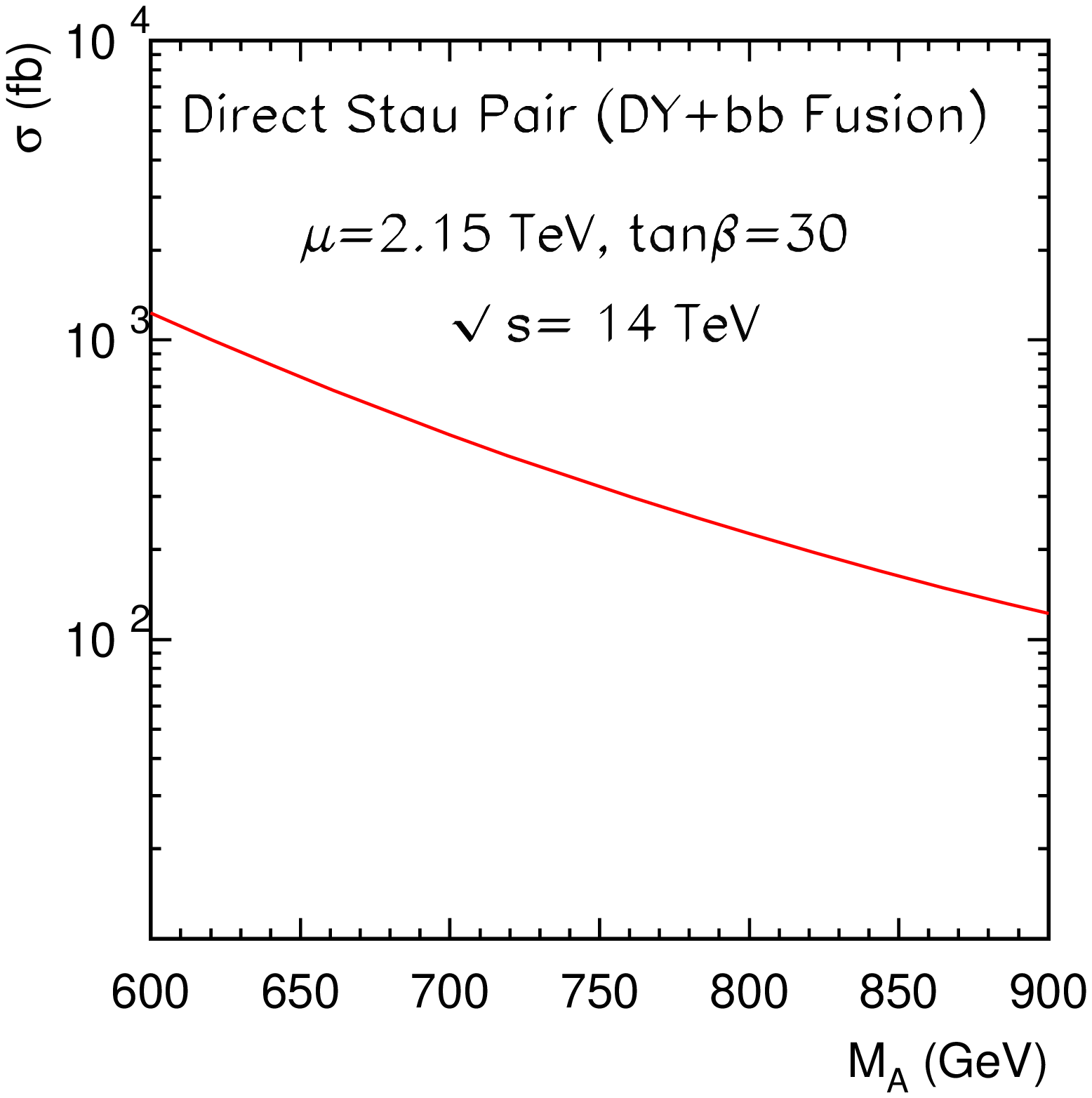}
\end{center}
\caption{(a) Gaugino pair production $\tilde{\chi}^{\pm}_{1}\tilde{\chi}^{0}_{2}$ (b) Direct stau pair including Drell-Yan and $b\bar{b}$ fusion} 
\label{rate}
\end{figure}
The significant enhancement from gaugino pair and $b\bar{b}$ fusion enables us to
search for such light staus in the early running of the LHC. 

\subsection{Like-sign Dilepton from $\tilde{\chi}^{\pm}_{1}\tilde{\chi}^{0}_{2}$}

\begin{figure}
\begin{center}

\includegraphics[scale=1,width=8cm]{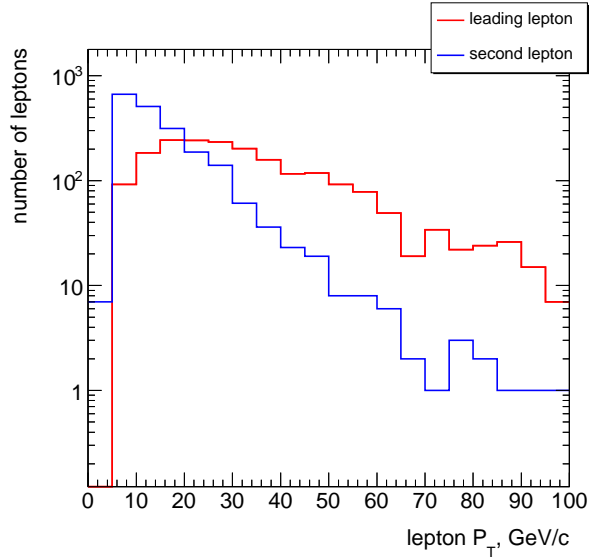}
\end{center}

\caption{$p_T$ of leptons from $\tau$ leptonic decay in like-sign dilepton sample due to $\tilde{\chi}^\pm_1\tilde{\chi}^0_2$ production at 7 TeV with $M_2=300$~GeV. The distribution is normalized by $20000/\sigma_{\tilde{\chi}^\pm_1\tilde{\chi}^0_2}$.} 
\label{pte}
\end{figure}

In the limit of $M_{2}=125$~GeV, stau pair from $\tilde{\chi}^{\pm}_{1}\tilde{\chi}^{0}_{2}$ and
$\tilde{\chi}^{+}_{1}\tilde{\chi}^{-}_{1}$ are over 3 pb which completely dominates the stau pair production. 
The stau $\tilde{\tau}^{\pm}_{1}$ further decays into $\tau^{\pm}$ and the dark matter candidate $\tilde{\chi}^{0}_{1}$ with 
\beq
\text{Br}(\tilde{\tau}^{\pm}_{1}\to \tau^{\pm} \tilde{\chi}^{0}_{1} )=100\%~. 
\eeq
$\tau^{\pm}$ final states have been playing important role in searching neutral Higgs
or charged Higgs Bosons. The pure leptonic $\tau^{\pm}$ decay is 
\beq
\text{Br}(\tau^{+}\to \ell^{+} \nu_{\ell}\bar{\nu}_{\tau})\simeq 35\%, \ell=e,\mu~.
\eeq
From detector perspective, the leptonic final states due to $\tau^{\pm}$
are indistinguishable from the direct lepton states. The searches of
leptonic taus then fall into the category of pure lepton searches. 
As mentioned,  50\% of $\tilde{\chi}^{\pm}_{1}\tilde{\chi}^{0}_{2}$ 
contribute to the same-sign stau pairs $\tilde{\tau}^{\pm}_{1}\tilde{\tau}^{\pm}_{1}$.
Leptons are from the $\tau^{\pm}$ three-body decay while $\tau^{\pm}$ from
$\tilde{\tau}^{\pm}_{1}$. In this scenario, the mass difference between stau states and LSP is 
\beq
\Delta m= m_{\tilde{\tau}_{1}}-m_{\tilde{\chi}^{0}_{1}} \simeq 35~\text{GeV}~.
\eeq
The other $\tau$ then depends on the mass difference between the neutralino $\tilde{\chi}^0_2$ and stau $\tilde{\tau}^\pm_1$. Since the off-diagonal terms in stau mass matrix are much larger than the diagonal, the $\tilde{\tau}^\pm_1$ states is a 50\% mixture state of $\tilde{\tau}_L$ and $\tilde{\tau}_R$. In the tau three-body decay, lepton is always boosted in the opposite direction to its spin. We use MadEvent \cite{madevent} with Pythia \cite{pythia} and TAUOLA \cite{tauola} and PGS-ATLAS for detector simulation. The lepton $p_T$ distribution which is normalized by $20000/\sigma_{\tilde{\chi}^\pm_1\tilde{\chi}^0_2}$ is given in Fig. \ref{pte}. 
The usual $p_T$ selection of lepton will further cut significant portion of the events.  

We employ the event selection following the criteria described in \cite{likesign}. 
They require the events to contain at least two electrons or muons of the same electric charge:
\begin{itemize}
\item $p_T > 20$~GeV, $\text{max}\{p_T\} > 25$~GeV for electron and 20~GeV for muon.
\item $\mid \eta_e \mid < 2.47$ with exclusion from $1.37 <\mid \eta_e \mid < 1.52$, $\mid \eta_\mu \mid < 2.5$
\item  Electrons and muons are required to be separated $\Delta R > 0.4$ from any jet with $p_T > 25~ \text{GeV} + 0.05 \times p_T(l)$.
\item $M_{\ell^\pm\ell^\pm}> 15~$GeV. Exclusion of the electron invariant mass $M_{e^\pm e^\pm}$ from 70 to 110~GeV to veto the $Z\to e^+ e^-$ events due to larger electron mis-charge ID rate.
\end{itemize}

Table \ref{likesigntable} shows the cut efficiencies for different benchmark points of $M_2$. 

\begin{table}
\begin{center}
\begin{tabular}{|c|c|c|c| c|}
\hline
$M_2$~(GeV) & 125 & 200 & 300 & 400 \\
\hline 
\hline
cut efficiency & 0.23\% & 0.26\% & 0.42\% & 0.57\%\\
\hline
\end{tabular}
\caption{Cut efficiency for inclusive like-sign dilepton} 
\end{center}
\label{likesigntable}
\end{table}%

With heavier $\chi^0_2$, the boost of $\tau$s are larger and 
the cut efficiency is then higher.

\begin{figure}
\begin{center}

\includegraphics[scale=1,width=8cm]{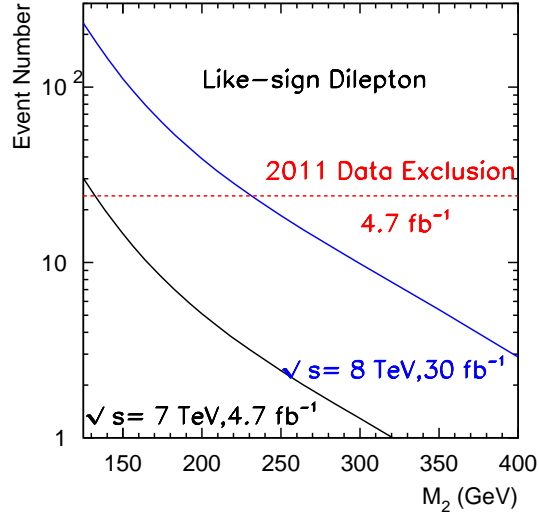}
\end{center}

\caption{Event number that passes through the same-sign dilepton selection cuts for varying $M_{2}$.} 
\label{pte2}
\end{figure}

The latest search result is from the 2011 data of 4.7~fb$^{-1}$ with 7 TeV running.
The maximal allowed same-sign dilepton events are 24 including both electrons
and muons events for $M_{\ell^\pm \ell^\pm}> 15$ GeV.
We then calculate the event numbers that can pass the selection cuts for like-sign dilepton search for
7 TeV LHC with 4.7~fb$^{-1}$ data and parameter region with $M_2< 132$~GeV has been excluded 
by the 2011-Data. Figure \ref{pte2} summarizes the event numbers of the Like-sign dilepton from
 $\tilde{\chi}^{\pm}_{1}\tilde{\chi}^{0}_{2}$ production and decay into stau final states. 
In order to estimate the discovery potential of 2011+2012 Data, we calculate the 
predicted events number of 30~fb$^{-1}$ Data for 8 TeV LHC which is slightly
higher than the 10~fb$^{-1}$ with 7 TeV running and 20~fb$^{-1}$ with 8 TeV running
to illustrate the qualitative feature. 

\subsection{Hadronic tau tag}
We then study searching $\tau^{\pm}$ through the hadronic $\tau$-tagging. 65\% of $\tau$ leptons
decay into hadronic final states. However, the $\tau$-jets are very different from the QCD jets in
its jet-energy shape. With less QCD activity, the $\tau$-jets are much narrow then the usual QCD jets.
Without loosing generality, we estimated hadronic $\tau$'s identification rate and the corresponding 
jet rejection rate following the performance presented in \cite{tauID}:
\begin{eqnarray}
& \eta_{\tau}=60\%, & R_{j}=5\%\nonumber\\
& \eta_{\tau}= 24\%,  & R_{j} = 1\%
\end{eqnarray}
The rejection rate $R_{j}$ corresponds to the faking rate of quark jet or gluon jet being misidentified
as a $\tau$-jet. Since we suffer from the small number of the events, we take the $\eta_{\tau}$ to be 60\% in the study.

In the inclusive stau pair production, the initial state radiation jet is kicked by the heavy particle pairs \cite{wangkai}.
It is useful to require a hard-jet in the final state. We require two $\tau$-tagged jets plus one hard jet and large missing transverse energy in the final state
\beq
j+ 2\tau_h +\cancel{E}_T~.
\eeq

The irreducible background is then $j+\tau^+ \tau^- +\nu\bar{\nu}$ which arises from $j+WW$ or $j+ZZ$ production.
However, the reducible background of jet being faked as tau turns out to be the largest background.
With one $\tau^{\pm}$ and $\cancel{E}_{T}$ in the final states, the leading reducible background is $Wjj\to \tau_h\nu+jj$ 
that was misidentified as $\tau_h\tau_h+j$. To suppress such background, we require the system transverse mass
$M_{T}$ to be larger than $m_{W}=80~$GeV. Since there were two $\tau$ in the final states, 
$M_{T}$ is defined by requiring the $\Delta\phi$ between the $\vec{p}^{\tau}_{T}$ and $\cancel{\vec{p}}_{T}$ is
the minimal one. 

We propose two searching mono-jet plus stau pairs ($j+ \tilde{\tau}_{1}^{+}\tilde{\tau}_{1}^{-}$).
The selection cuts are designed as 
\begin{itemize}
\item the hardest jet with $p_{T}> 80$~GeV
\item $\cancel{E}_{T}> 100$~GeV
\item identification of at least two $\tau_h$ (reconstructed $p_{T}^{\tau}>25$~GeV)
\item $M_{T}>80$~GeV ($M_{T}$ defined by minimum $\Delta\phi$ between $\vec{p}^{\tau}_{T}$ and $\cancel{\vec{p}}_{T}$)
\end{itemize}
The multi-jet plus $Z\to \nu\bar{\nu}$ final state can also contribute as reducible background. 
In addition, even the pure multijet final states can contribute as reducible background. The pure multijet final states
may also have missing transverse energy due to mis-measurement of jet energy, detector crack, or even neutrinos
from the meson decays if it's not corrected to the jet. Therefore, we require a large $\cancel{E}_T>100$~GeV to
suppress such background.
The backgrounds in Table \ref{table} are estimated by irreducible background of $j+\tau^{+}\tau^{-}+\cancel{E}_{T}$
and the case where one jet or two jets were misidentified as $\tau$-jet using the above jet rejection rate $R_{j}$. 
\begin{table}
\label{table}
\begin{center}

\begin{tabular}{|c|c|c|c|c|}
\hline 
  & $j+\tau^{+}\tau^{-}+\cancel{E}_{T}$ & $2j+\tau^{\pm}+\cancel{E}_{T}$ & $3j+\nu\bar{\nu}$ & 3j \tabularnewline
\hline 
\hline 
$\sigma$(pb)  & 0.34 & 1280 & 670 & $7.8\times 10^7$ \tabularnewline
\hline 
$p^{j}_{T}\geqslant$80~GeV  & 29.67\% & 20.53\% & 42.02\% & 7.75\% \tabularnewline
\hline 
$\cancel{E}_{T}\geqslant$100~GeV  & 24\% & 6.5\% & 22\% & $< 10^{-5}$\tabularnewline
\hline 
$\mathrm{N}_{\tau}\geqslant2$ ($\eta_{\tau}=60\%$) & 10.14\% & 1.3\% & 0.22\% & 0.12\% \tabularnewline
\hline 
$M_{T}\geqslant$80~GeV & 16.27\% & 4.84\% & 42.70\% & $< 10^{-5}$\tabularnewline
\hline 
$\sigma_{\mathrm{cut}}$(fb)   & 0.4 & 10.6 & 59.0 & $<10^{-3}$  \tabularnewline
\hline 
\end{tabular}
\end{center}

\caption{Cut efficiency for SM irreducible and reducible background for one hard-jet plus two tagged hadronic $\tau^\pm$s. }
\end{table}
The pure multijet final states are several orders higher to begin with and even with $\cancel{E}_T>100$~GeV cut, 
it is still the leading background. Fortunately, the $M_T>80$~GeV cut also significantly reduce the background
of this kind. If the large $\cancel{E}_T$ is due to mis-measurement or neutrinos from jets, the $M_T$ only reconstruct the jet mass which is typically small. Therefore, the $M_T$ cut is also very efficient for cutting pure multijet background. 

We then calculate the signal cut efficiency for different benchmark points in Table \ref{m2} and Table \ref{ma}.
\begin{table}
\begin{center}

\begin{tabular}{|c|c|c|c|c|}
\hline 
$M_{2}\mathrm{(GeV)}$ & 125 & 200 & 300 & 400\tabularnewline
\hline 
\hline 

$p_{T}^{j}\geqslant80\mathrm{GeV}$ & 30.24\% & 37.48\% & 48.22\% & 54.65\%\tabularnewline
\hline 
$\cancel{E}_{T}\geqslant100\mathrm{GeV}$ & 22.47\% & 31.66\% & 45.80\% & 56.56\%\tabularnewline
\hline 
$N_{\tau}\geqslant2,(\eta_{\tau}=60\%)$ & 8.24\% & 17.20\% & 19.85\% & 21.28\%\tabularnewline
\hline 
$M_{T}\geqslant80\mathrm{GeV}$ & 14.20\% & 26.80\% & 39.24\% & 48.07\%\tabularnewline
\hline 
 
\end{tabular}
\end{center}

\caption{Cut efficiency for signal benchmarks by varying $M_{2}\mathrm{(GeV)}$}
\label{m2}
\end{table}

\begin{table}
\begin{center}

\begin{tabular}{|c|c|c|c|c|}
\hline 
$M_{2}\mathrm{(GeV)}$ & 600 & 700 & 800 & 900\tabularnewline
\hline 
\hline 

$p_{T}^{j}\geqslant80\mathrm{GeV}$ & 34.73\% & 37.96\% & 39.85\% & 39.89\%\tabularnewline
\hline 
$\cancel{E}_{T}\geqslant100\mathrm{GeV}$ & 24.60\% & 28.29\% & 30.82\% & 31.24\%\tabularnewline
\hline 
$N_{\tau}\geqslant2,(\eta_{\tau}=60\%)$ & 12.21\% & 12.42\% & 12.89\% & 12.70\%\tabularnewline
\hline 
$M_{T}\geqslant80\mathrm{GeV}$ & 19.16\% & 22.22\% & 23.74\% & 26.22\%\tabularnewline
\hline 

\end{tabular}\caption{Cut efficiency for signal benchmarks by varying $M_{A}\mathrm{(GeV)}$}
\end{center}

\label{ma}
\end{table}

\begin{figure}
\begin{center}

\includegraphics[scale=1,width=8cm]{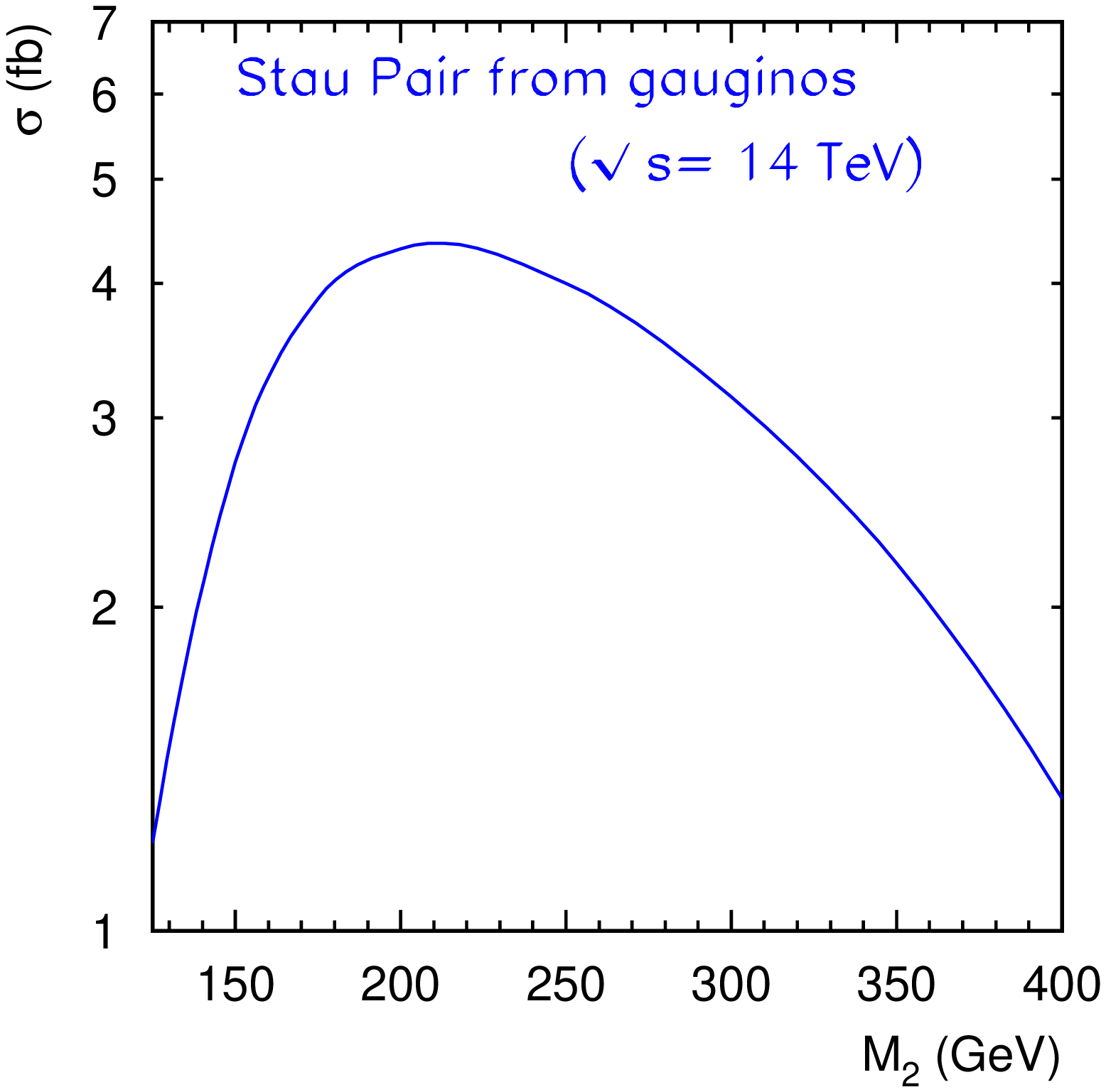}
\includegraphics[scale=1,width=8cm]{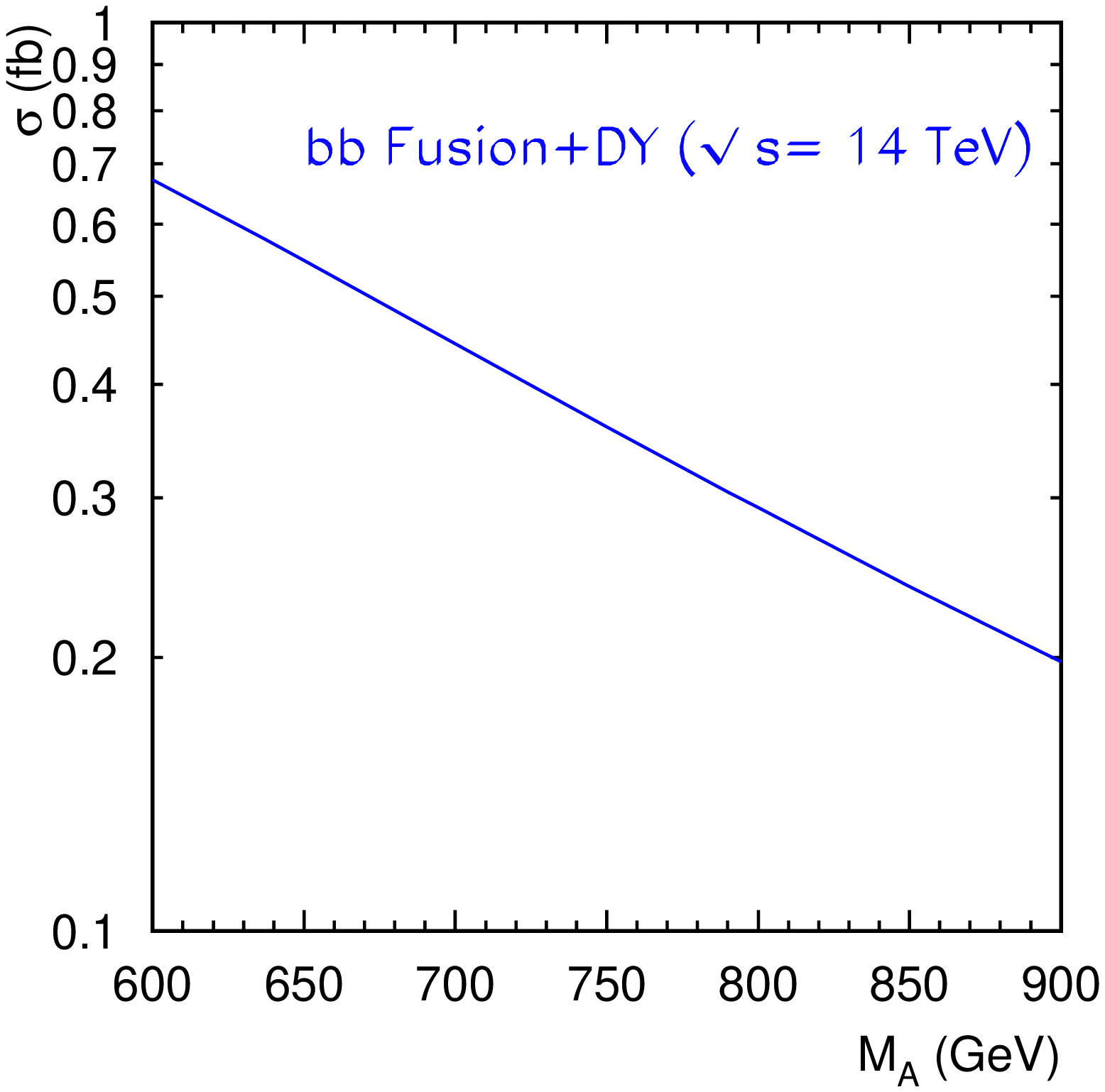}
\end{center}

\caption{(a) Gaugino pair production $\tilde{\chi}^{\pm}_{1}\tilde{\chi}^{0}_{2}$ (b) Direct stau pair including Drell-Yan and $b\bar{b}$ fusion. All the results are after cuts.} 
\label{rate3}
\end{figure}

Table \ref{m2} clearly shows that the cut efficiency is higher with larger $M_2$ due to larger mass difference. 
Therefore, the total cross section maximizes at around $M_2\simeq 220$~GeV as shown in Fig. \ref{rate3}

\begin{figure}
\begin{center}

\includegraphics[scale=1,width=8cm]{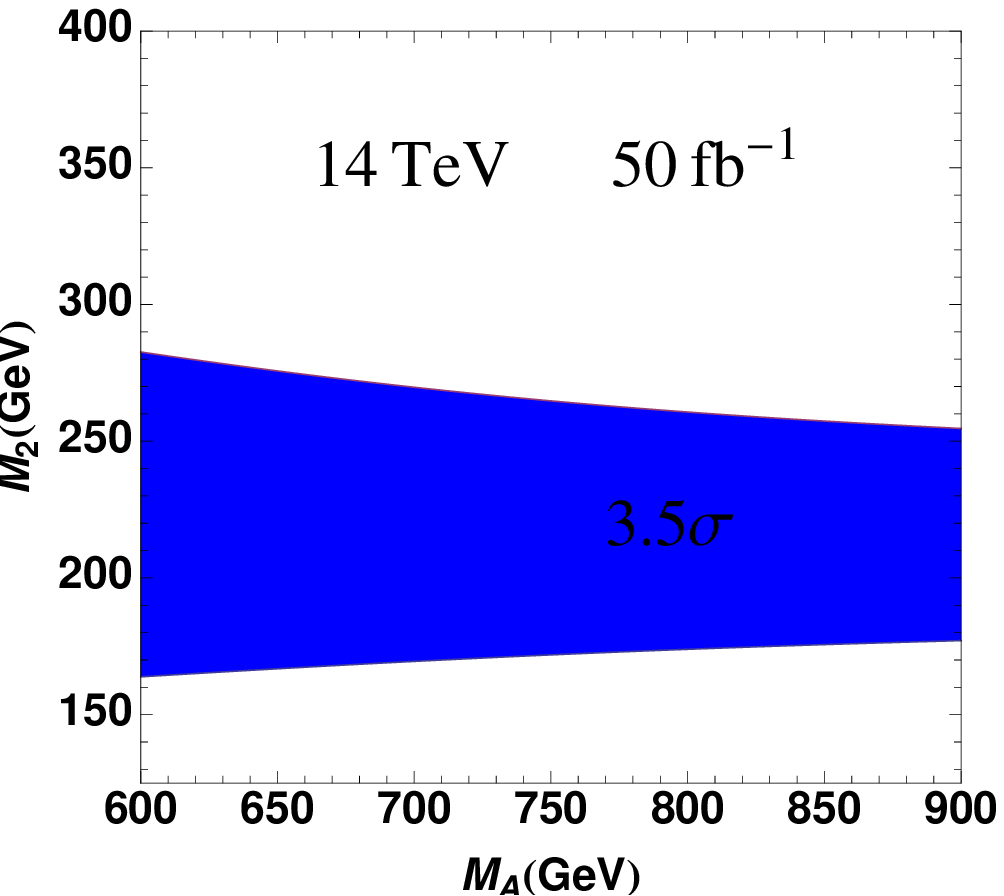}
\includegraphics[scale=1,width=8cm]{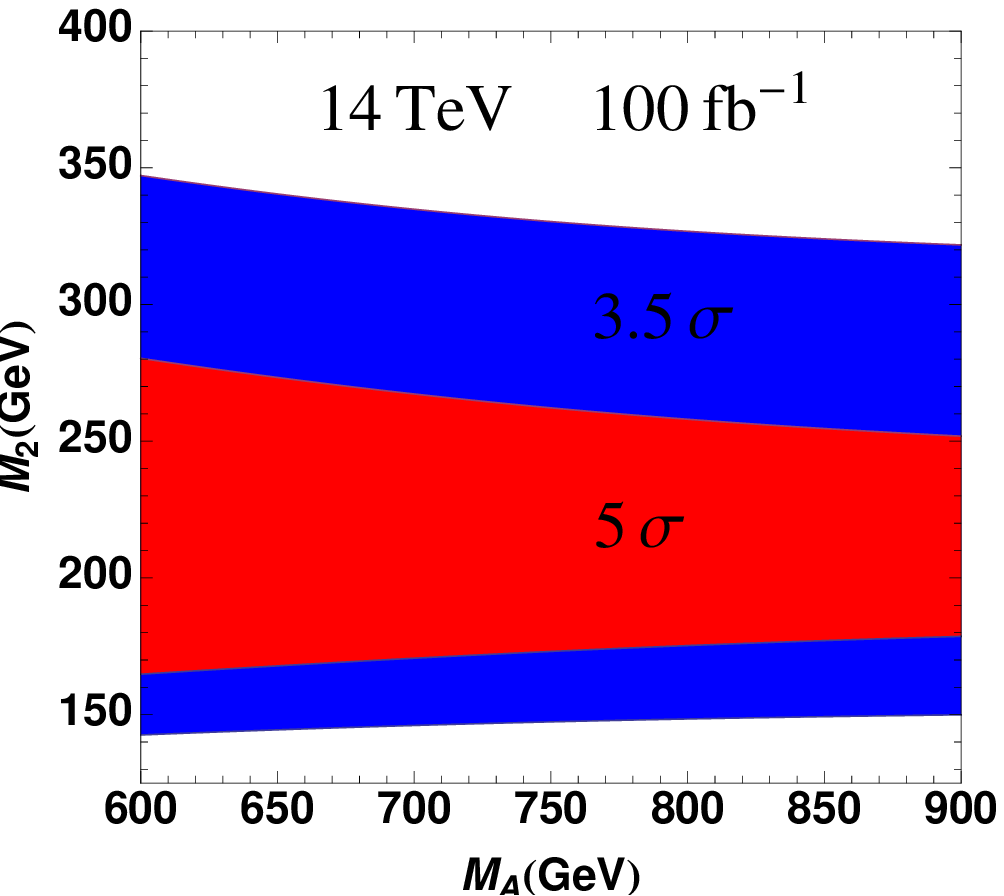}
\end{center}

\caption{Contours for 3.5$\sigma$ and 5$\sigma$ significance for stau search through hadronic tau tagging. } 
\label{contour}
\end{figure}

We summarize discovery potential for 14 TeV LHC with 50 and 100 fb$^{-1}$ data of inclusive stau pair in Fig. \ref{contour}

Even with the most optimistic region of $M_2\sim 200-300$~GeV, it requires at least 50~fb$^{-1}$ data of 14 TeV LHC to reach a significance of 3.5 $\sigma$. Then we conclude it is
difficult to claim discovery only through hadronic tau based on the data by the 2012 shut-down. 5 $\sigma$ reach for our most optimistic region then require 100~fb$^{-1}$ data  with 14 TeV running.

\section{Conclusions}
In order to explain the enhanced excess of the di-photon signal in Higgs search, $\sigma_{\gamma\gamma}/\sigma_{\rm SM}\simeq 1.5$, we argue that the models with enhanced $\Gamma(h\to \gamma\gamma)$ alone is the most favorable scenario when combing the latest LHC and
the Tevatron results. To illustrate this feature, we discuss a scenario within supersymmetry framework with light stau that predicts a 125 GeV SM-like Higgs boson and is in consistent with all the constraints including dark matter and $B_{s}\to \mu^{+}\mu^{-}$ etc \cite{carlos, liantao}.
we focus on the parameter space with enhanced production of inclusive stau pairs, 
in particular, via $b\bar{b}$ fusion or gaugino pairs. We study its phenomenology in both pure leptonic tau $\tau^\pm_\ell$ channels and hadronic tau tagged $\tau_h$ channels. We find the same-sign dilepton from $\tilde{\chi}^\pm_1 \tilde{\chi}^0_2 \to \tau^\pm_\ell \tau^\pm_\ell+X$ may significantly improve the discovery potential with even 7--8 TeV LHC of ${\cal O}(30~\text{fb}^{-1})$ data. In the case of hadronic tau pair, we use the final state $j+\tau_h \tau_h +\cancel{E}_T$ to search
and find that even with the most optimistic region of $M_2\sim 200-300$~GeV, it requires at least 50~fb$^{-1}$ data of 14 TeV LHC to reach a significance of 3.5 $\sigma$. We conclude it is
difficult to claim discovery only through hadronic tau based on the data by the 2012 shut-down. 5 $\sigma$ reach for our most optimistic region then require 100~fb$^{-1}$ data  with 14 TeV running and one will have to rely the same-sign dilepton search.  

\section*{Acknowledgement}
We would like to thank Shan Jin and Xuai Zhuang from China ATLAS collaboration 
for useful discussion. ML is supported by the National Science Foundation of China (10875103, 11135006), National Basic Research Program of China (2010CB833000), and Zhejiang University Group Funding (2009QNA3015). LS is supported by the National Science Foundation of China (11061140513) under China ATLAS collaboration. KW'S work is supported in part, by the Zhejiang University Fundamental Research Funds for the Central Universities 2011QNA3017 and the National Science Foundation of China (11245002,11275168).

\end{document}